\documentclass{article}
\usepackage[utf8]{inputenc}
\usepackage{bm} 
\usepackage{amsmath}
\usepackage{authblk}
\usepackage{amssymb}
\usepackage{graphicx}
\usepackage{hyperref}
\usepackage{tikz}
\usepackage{subcaption}
\usepackage{color}

\usepackage{marginnote}

\usepackage[
per-mode=fraction,
separate-uncertainty,
retain-unity-mantissa=false,
product-units=power,
table-alignment=center,
detect-all,
binary-units=true,
exponent-base=10
]{siunitx}

\newlength\Colsep
\newlength{\savedparindent}

\title{Simultaneous Multi-Slice MRI using Cartesian and Radial FLASH 
and Regularized Nonlinear Inversion: SMS-NLINV}
\author{Sebastian Rosenzweig$^{1}$, H. Christian M. Holme$^{1,2}$, Robin N. 
Wilke$^{1,2}$, Dirk Voit$^{3}$, Jens Frahm$^{2,3}$ and Martin Uecker$^{1,2}$}

\affil{
	$^1$Institute for Diagnostic and Interventional Radiology,\\
	University Medical Center Göttingen, Göttingen, Germany.\\
	$^2$German Centre for Cardiovascular Research (DZHK),\\
	Partner site Göttingen, Göttingen, Germany.\\
	$^3$Biomedizinische NMR Forschungs GmbH am 
	Max-Planck-Institut für biophysikalische Chemie, Göttingen, Germany.}

\begin{document}
\setlength{\savedparindent}{\parindent}

\maketitle

\noindent
{\em Running head:} SMS Reconstruction Using Nonlinear Inversion

\noindent
{\em Address correspondence to:} \\
Sebastian Rosenzweig\\
University Medical Center Göttingen \\
Institute for Diagnostic and Interventional Radiology \\
Robert-Koch-Str. 40\\
37075 Göttingen, Germany \\
sebastian.rosenzweig@med.uni-goettingen.de

\vspace{0.3cm}
\noindent
Approximate word count: 153 (Abstract) 3470 (body)

\vspace{0.3cm}
\noindent
Submitted to {\it Magnetic Resonance in Medicine} as a Full Paper.

\vspace{0.3cm}
\noindent
Part of this work has been presented at the ISMRM Annual Conference 2016 
(Singapore) and 2017 (Honolulu).
This work was supported by a seed grant of the	
Physics-to-Medicine Initiative Göttingen (LM der Niedersächsischen Vorab).

\newpage
\setlength{\parindent}{0in}
\section*{Abstract}
{\bf Purpose:}
The development of a calibrationless parallel imaging
method for accelerated simultaneous multi-slice (SMS)
MRI based
on Regularized Nonlinear Inversion (NLINV), evaluated using Cartesian and 
radial FLASH.

{\bf Theory and Methods:}
NLINV is a parallel imaging method that jointly estimates image content
and coil sensitivities using a Newton-type method with regularization.
Here, NLINV is extended to SMS-NLINV for reconstruction and separation
of all simultaneously acquired slices.
{The performance of the
extended method is evaluated for different sampling schemes using
phantom and in-vivo experiments based on Cartesian and radial SMS-FLASH
sequences.

{\bf Results:}
The basic algorithm was validated in Cartesian experiments by
comparison with ESPIRiT. For Cartesian and radial sampling, improved results
are demonstrated compared to single-slice experiments, and
it is further shown that sampling schemes using complementary samples
outperform schemes with the same samples in each partition.

{\bf Conclusion:}
The extension of the NLINV algorithm for SMS data was
implemented and successfully demonstrated in
combination with a Cartesian and radial SMS-FLASH sequence.

\vspace{0.5in}
{\bf Key words: simultaneous multi-slice, SMS, multi-band, regularized 
nonlinear inversion, NLINV, parallel  
imaging}
\newpage

\setlength{\parindent}{\savedparindent}
\section*{Introduction}

Accelerating image acquisition is of great importance in clinical
magnetic resonance imaging (MRI). Parallel imaging
exploits receive-coil arrays for acceleration.
Conventional reconstruction methods for parallel imaging consist
of a calibration from reference lines followed
by linear reconstruction \cite{Pruessmann_Magn.Reson.Med._1999,
Griswold_Magn.Reson.Med._2002, Lustig_Magn.Reson.Med._2010,
Uecker_Magn.Reson.Med._2014}. In contrast, Regularized Nonlinear Inversion (NLINV)
\cite{Uecker_Magn.Reson.Med._2008} does not require a calibration step
but simultaneously computes image content and coil sensitivities from all
available data.
Because NLINV does not depend on the presence of explicit (Cartesian) 
calibration data,
it is ideally suited for non-Cartesian parallel imaging. For example,
NLINV is used in a highly accelerated real-time MRI method based on
radial sampling \cite{Uecker_NMRBiomed._2010}.

Many applications require the acquisition of several slices. 
Simultaneous multi-slice (SMS) MRI \cite{Larkman_J.Magn.Reson.Imaging_2001}
allows for significant scan time reductions and improved image
quality \cite{Moeller_Magn.Reson.Med._2010, Feinberg_PLoSOne_2010}. In SMS
MRI several slices are excited at the same time and the resulting
superposition is disentangled using special encoding schemes
\cite{Maudsley_J.Magn.Reson._1980, Souza_J.Comput.Assist.Tomogr._1988}
and/or the spatial encoding information inherent in receiver coil
arrays \cite{Larkman_J.Magn.Reson.Imaging_2001}.
The main benefit of accelerated SMS MRI over conventional single-slice
imaging is the possibility to distribute undersampling among an additional dimension
and exploit sensitivity encoding in all three dimensions which allows for higher
acceleration factors \cite{Weiger_Magn.Reson.Mater.Phys.Biol.Med._2002,
Wiesinger_Magn.Reson.Med._2004,Breuer_Magn.Reson.Med._2005,Yutzy_Magn.Reson.Med._2011,
 Wang_Magn.Reson.Imaging_2016,
Setsompop_Magn.Reson.Med._2012}.

The aim of this work is to extend NLINV for the reconstruction of
SMS data. First, the extension of the algorithm for Cartesian
and radial sampling with arbitray encoding in slice direction
is introduced. For Cartesian data from an SMS-FLASH sequence,
SMS-NLINV is compared to ESPIRiT \cite{Uecker_Magn.Reson.Med._2014}.
For Cartesian and radial data,
a single-slice measurement is compared to SMS acquisitions with equivalent or 
complementary samples in each partition. Accelerated SMS measurements
of a human brain and a human heart are performed to show feasibility of in vivo 
scans.

\section*{Theory}
Table \ref{Tab:Notation} shows the notation used in this work.

\subsection*{SMS Encoding and Excitation Pulses}
	\label{SS:Fourier-encoding and RF design}
	In SMS MRI, $M$ partitions $p=1,\dots,M$ are measured to 
	get 
	information about $M$ parallel slices $q=1,\dots,M$. 
	Please note that a fully sampled acquisition with 
	$M$ partitions has $M$ times the number of samples compared to a 
	single-slice experiment, and the acceleration factor of an SMS experiment 
	is then 
	given by $R=N^\text{full}/N^\text{red}$, with $N^\text{full/red}$ the 
	number 	of samples acquired in a full and undersampled partition 
	measurement, 
	respectively. Contrary to 
	conventional 
	multi-slice, in each partition measurement all $M$ slices are excited 
	simultaneously, 
	i.e.\ superposed data are acquired.
	In the limit of small flip angles, an SMS radio frequency (RF) 
	excitation
	pulse which excites $M$ slices at positions $z_q$ and
	thickness $\Delta z_q$ can be created by superposing
	conventional single-slice excitation pulses $B_\text{rf}^{(1)}(z_q,\Delta 
	z_q)$.
	To generate differently encoded partitions a 
	unitary $M\times M$ encoding matrix $\Xi$ is included.
	The SMS RF excitation pulse for partition measurement $p$ is then given by
	\begin{equation}
	\label{Eq:SMS pulse}
	\tilde{B}_{\text{rf},p}^{(M)}(z_1,\dots,z_M,\Delta z_1,\dots,\Delta z_M) :=
	\sum_{q=1}^M \Xi_{pq} B_\text{rf}^{(1)}(z_q,\Delta z_q).
	\end{equation}
	Let $\bm{y}_q:=(y_q^1,\dots,y_q^N)$ be a vector which contains the k-spaces
	$y_q^j$ of slice $q$ and coils $j=1,\dots,N$. Then, the encoded
	k-space of partition $p$ is given by
	\begin{equation}
	\label{Eq:k encoded}
	\tilde{\bm{y}}_p := \sum_{q=1}^M \Xi_{pq} \bm{y}_q.
	\end{equation}
	Although the derived SMS-NLINV algorithm is completely generic, we use the
	discrete Fourier-matrix for encoding
	in the scope of this work, i.e.
	\begin{equation}
	\label{Eq:DFT matrix}
	\Xi_{pq} = \exp\left(-2\pi i \frac{(p-1)(q-1)}{M}\right),\;\; 
	p,q=1,\dots,M.
	\end{equation}

\subsection*{Image reconstruction}
	If the encoded k-spaces
	$\tilde{\bm{y}}_1,\dots,\tilde{\bm{y}}_M$ determined by the $M$
	partition measurements are
	fully sampled, the k-space of each slice can be recovered by applying the
	inverse of the encoding matrix
	\begin{equation}
	\label{Eq:k encoded inverse}
	{\bm{y}}^\text{avg}_q := \sum_{p=1}^M \Xi^{-1}_{qp} \tilde{\bm{y}}_p.
	\end{equation}
        Note that the k-spaces $\bm{y}_q^\text{avg}$ possess an SNR benefit of
	$\sqrt{M}$ compared to single-slice experiments due to averaging given by
	Eq.\ (\ref{Eq:k encoded inverse}) and because $\Xi$ is unitary. 
	Eq.\ (\ref{Eq:k encoded inverse}) can also be applied to undersampled data
	if the same k-space positions are acquired for all partitions. The recovered
	(but still undersampled) k-spaces
	$\bm{y}^\text{avg}_q$ can then be processed using conventional single-slice
	reconstruction algorithms. This still leads to an SNR benefit, but the
	actual advantage of SMS - the acceleration in direction perpendicular
	to the slices - only comes into play when distinct samples are acquired
	for each partition. Then,
	Eq.\ (\ref{Eq:k encoded inverse}) is no longer applicable and more
	elaborate SMS reconstruction approaches must be applied.
	A novel approach to tackle this
		reconstruction problem is introduced in the following.

	Regularized Nonlinear Inversion (NLINV)
	\cite{Uecker_Magn.Reson.Med._2008} can be extended for the
	reconstruction of encoded SMS data \cite{Rosenzweig__2016}. In NLINV, the 
	MRI signal equation is
	modeled as a nonlinear operator equation,
	\begin{equation}
		F({X})=\tilde{Y}.
		\label{Eq:Signal}
	\end{equation}
	$X$ is the vector to be reconstructed. It contains the image content
	$m_q(\bm{r})$ and the $N$ coil sensitivities $c^{j}_q(\bm{r}),\;
	j=1,\dots,N$,  for each of the $M$ slices $q$, i.e.\ the stacked vector
	$X := \left( \bm{x}_1, \dots, \bm{x}_M \right)^T$ as a
	concatenation of the vectors $\bm{x}_q := \left( m_{q}, c_q^1 \dots, c_q^N 
	\right)^T$.
	The vector $\tilde{Y}$ contains the encoded k-spaces for all $M$ 
	partitions and all $N$ channels, i.e.\
	$\tilde{Y}:=\left(\tilde{\bm{y}}_{1},\dots,\tilde{\bm{y}}_{M}\right)^T$
	with 
	$\tilde{\bm{y}}_{p}:=\left(\tilde{y}_{p}^1,\dots,\tilde{y}_{p}^N,\right)^T$.
	Then, the nonlinear mapping function $F$ is given by
	\begin{equation}
	\label{Eq:Forward op}
		F:{X}\mapsto 
		\bm{P}
		\Xi
		\left(\begin{array}{c}
			\mathcal{F} (m_1 \bm{c}_1) \\
			\vdots \\
			\mathcal{F} (m_{M} \bm{c}_{M})
		\end{array}\right),\;\;
		\mathcal{F} (m_q \bm{c}_{q}) :=
		\left(\begin{array}{c}
		\mathcal{F} (m_q c_q^1) \\
		\vdots\\
		\mathcal{F} (m_q c_q^N)
		\end{array}\right).
	\end{equation}
	Here, $\mathcal{F}$ is the (two-dimensional) Fourier transform and $\Xi$ is 
	an encoding matrix, e.g.\ Eq.\ (\ref{Eq:DFT matrix}). The projection 
	matrix  $\bm{P}$ is defined by
	\begin{equation}
	\label{Eq:Projector}
	\bm{P} := \left(\begin{array}{ccc}
	P_1 & & 0 \\
	& \ddots & \\
	0 & & P_{M}
	\end{array}\right),
	\end{equation}
	where ${P}_p$ is the orthogonal projection onto the k-space trajectory used 
	for partition $p=1,\dots,M$.
	A more compact notation for Eq.\ 
	(\ref{Eq:Forward op}) can be given by 
	introducing the operator $\mathcal{C}$, which performs the 
	multiplication of the object with the sensitivities:
	\begin{equation}
	\label{Eq:Forward op compact}
	F: X \mapsto \bm{P}\Xi\mathcal{F} \mathcal{C} X.
	\end{equation}
	The forward operator $F$ weights the magnetization $m_q$  of slice $q$ 
	with the coil sensitivities $\bm{c}_q=(c_q^1,\dots, c_q^N)^T$ ($\mathcal{C}$),
	transforms into k-space ($\mathcal{F}$), encodes 
	($\Xi$) and samples ($\bm{P}$).
	The derivative $DF$ and its adjoint $DF^H$ will be 
	used later to solve 
	the inverse problem Eq.\ (\ref{Eq:Signal}) and are given in the Appendix.
	Figure \ref{Fig:Flow Chart} shows a flow chart of the operators $F$,
	$DF$ and $DF^H$. 

	Eq.\ (\ref{Eq:Signal}) is highly 
	underdetermined, hence prior knowledge has to be incorporated to prevent 
	image content to be assigned to coil profiles and vice versa. While the 
	image content can contain strong variations and edges, coil profiles
	in general are smooth functions, so a smoothness-demanding norm 
	can be applied. Uecker et al.\ suggest a Sobolev norm
	\begin{equation}
	||f||_{H^{l}}:=||a(I-b\Delta)^{l/2}f||_{L^2},
	\label{Eq:Sobolev}
	\end{equation}
	with $l$ a positive integer, $I$ the identity matrix, $a$ and $b$ scaling 
	parameters and 
	$\Delta=\partial_x^2 + \partial_y^2$ the 
	2D 
	Laplacian. Hence, in Fourier 
	space the standard L2-norm has to be weighted by the additional term 
	$a(1+b||k||^2)^{l/2}$, which 
	penalizes high spatial frequencies. This regularization is implemented by 
	transforming $X=(\bm{x}_1,\dots,\bm{x}_M)^T$ using a weighting
	matrix $\mathcal{W}^{-1}$.
	We denote $X' := \mathcal{W}^{-1} X$ and
	$\bm{x}'_q := W^{-1} \bm{x}_q = \left(m_q, {c'}_q^1, \dots, {c'}_q^N, \right)^T$.
	This yields a transformed but equivalent system of equations
	\begin{equation}
	G(X') :=  F\mathcal{W} X' = \tilde{Y},
		\label{Eq:Signal2}
	\end{equation}
	which is solved using the Iteratively Regularized Gauss Newton Method 
	(IRGNM). 
	
	As a first step, the IRGNM 
	linearizes Eq.\ (\ref{Eq:Signal2}),
	\begin{equation}
	\tilde{Y}=DG\big|_{X'_n}dX' + GX'_n,
	\end{equation}
	where $X'_n$ is the estimate of the $n^\text{th}$ Newton step and 
	$DG\big|_{X'_n}$ 
	is the Jacobian of $G$ at $X'_n$. This equation is solved in the
	least-squares sense and with regularization using the 
	Conjugate Gradient algorithm. 
	The corresponding cost 
	function to be 
	minimized in every Newton step is
	\begin{equation}
	\Phi(dX') = \underset{dX'}{\text{argmin}}\left(||DG\big|_{X'_n}d{X'} - 
	(\tilde{Y}-GX'_n)||^2_{L^2} + \beta_n ||X'_n 
	+ d{X'}||^2_{L^2} \right),
	\label{Eq:Cost Function}
	\end{equation}
	where the L2 penalty term 
	$\beta_n||X'_n+dX'||^2_{L^2} =\beta_n||\mathcal{W}^{-1}X_{n+1}||^2_{L^2} 
	$, with 
	$\beta_n=\beta_0 h^n$ and $h\in(0,1)$, implies Tikhonov regularization.
	\begin{figure}[h]
		\centering
		\includegraphics[width=0.5\textwidth]{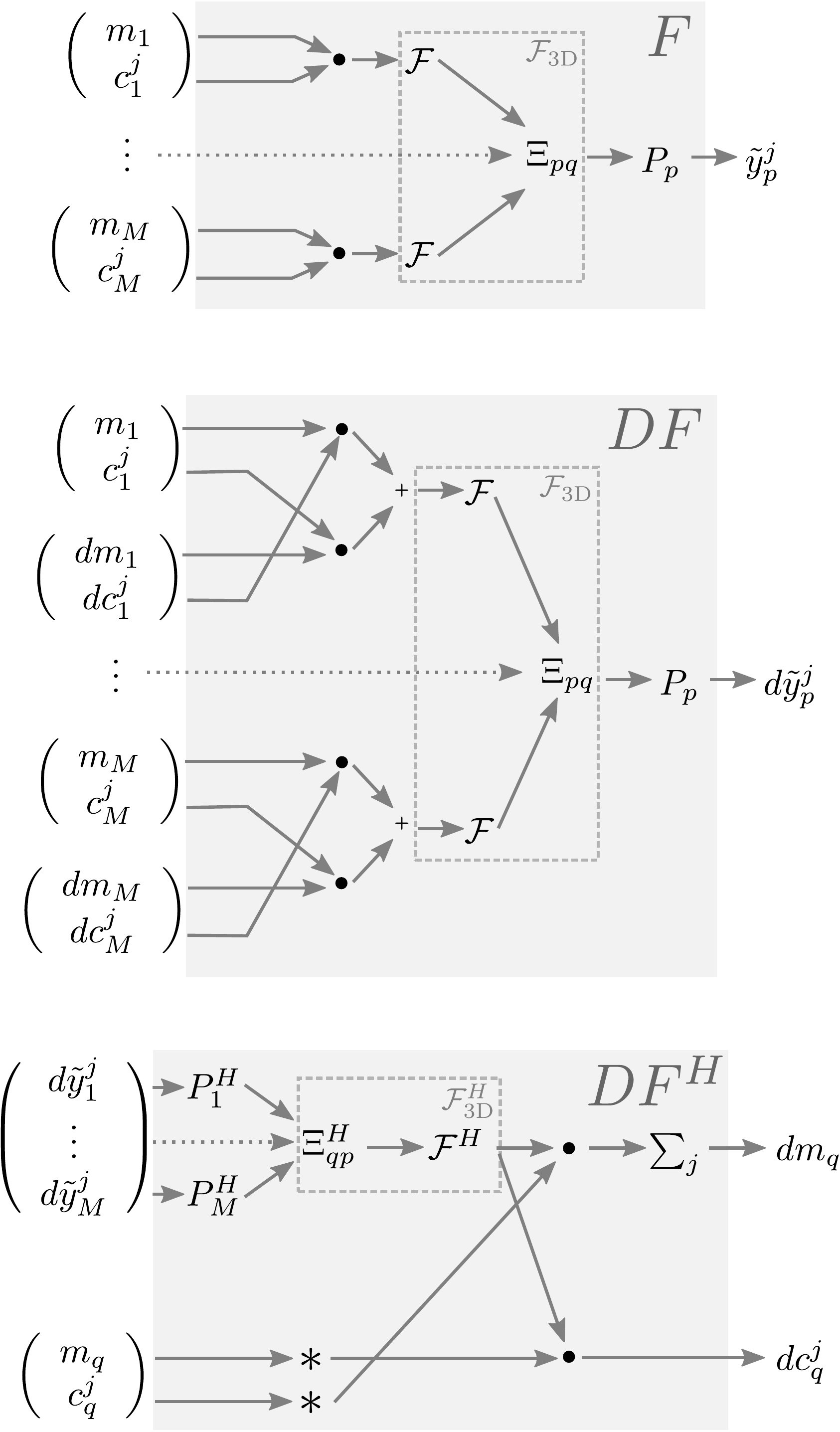}
		\caption{Flow chart for the calculation of the forward operator 
		$F$, its 	derivative $DF$ and the adjoint of the 
		derivative 
		$DF^H$. $\tilde{y}$: Encoded k-space data. $m$: 
		Magnetization. $c$: Coil sensitivity. 
		$P$: Projection onto k-space trajectory. $\mathcal{F}$: 2D Fourier 
		transform. $\Xi_{pq}$: Encoding matrix.
		$\bm{\cdot}$: 
		Pointwise 
		multiplication. 
		$+$: Addition. $*$: Complex conjugation.}
		\label{Fig:Flow Chart}
	\end{figure}

\subsection*{Implementation for Cartesian and non-Cartesian Data}

		We assume Eq.\ (\ref{Eq:Signal}) to be
	given in discretized form and all functions are represented by vectors of 
	point values on a rectangular grid. 
	For Cartesian sampling, $\mathcal{F}$
	can be implemented exactly as a discrete Fourier transform, 
	and ${P}_p$ is a 
	diagonal matrix with ones at sample positions and zeros elsewhere. 
	The 2D Fourier
	transform $\mathcal{F}$ always appears in combination with the
	encoding matrix $\Xi$, which in this work is the discrete Fourier-matrix
	Eq.\ (\ref{Eq:DFT matrix}). Thus, $\Xi \mathcal{F}$ and its adjoint can
	simply be implemented as a three-dimensional Fast
	Fourier transform and its adjoint.
	Note, that the 2D Fourier transform  $\mathcal{F}$ is a
	discretized version of a continuous Fourier transform, whereas the
	Fourier-encoding $\Xi$ is discrete by definition.

	For non-Cartesian sampling, $P_p$ projects onto arbitrary positions 
	in k-space. As in non-Cartesian SENSE, $P \mathcal{F}$ can be
	implemented with a non-uniform Fourier transform \cite{Knoll__2009}.
	The term $\mathcal{F}^H\Xi^H\bm{P}\Xi\mathcal{F}$
	is the main operation which occurs in each iteration step.
	As described previously 
	\cite{Wajer__2001},
	it can be interpreted as a non-periodic convolution with a point-spread function (PSF) which has to be evaluated
	on a region with  compact support defined by the field-of-view. Thus, an efficient implementation is achieved 
	with the fast Fourier transform on a $2$-fold enlarged grid to implement
	the non-periodic convolution using Toeplitz embedding. This requires only
	a minor modification of the Cartesian implementation which can then be used 
	with data
	gridded	once onto the Cartesian grid in a preparatory step and with a 
	pre-computed PSF.

\subsection*{Sampling Schemes}

All utilized Cartesian sampling patterns
possess  $L_\text{ref}$ 
reference lines in the k-space center, whereas the periphery is undersampled 
by a factor $R$. For each of the $M$ partition measurements we can use a 
distinct 
undersampling pattern. The CAIPIRINHA technique can improve the image quality
for SMS  acquisitions by 
acquiring alternating lines between each partition 
\cite{Breuer_Magn.Reson.Med._2005,Zhu__2012}. Alternatively, in 
each partition the 
same samples can be acquired (aligned pattern).

In radial measurements k-space samples are acquired along spokes. Let 
$N_\text{sp}$ be the total number of acquired spokes per partition. 
Then, the 
angle between consecutive spokes of a partition is set to
$\alpha_\text{sp}={2\pi}/N_\text{sp}$ which 
guarantees uniform 
k-space coverage and prevents strong gradient delay artifacts by opposing the 
acquisition direction of adjacent spokes. For each partition 
the k-space 
trajectory, i.e.\ the spoke distribution scheme, can be chosen individually. 
Figure \ref{Fig:Turn_scheme} shows three possible spoke distribution schemes:
(1) The aligned scheme acquires the same spokes for each partition.
(2) In the 
linear-turn scheme the initial spoke pattern is rotated by 
\begin{equation}
\alpha^\text{LIN}_\text{trn}=(p-1)\cdot \frac{\pi}{N_\text{sp}\cdot 
M}
\end{equation}
for partition $p=1,\dots,M$, assuring the acquisition of complementary 
samples and uniform 
spoke distribution (cf.\ Fig.\ \ref{Fig:Turn_scheme}b). 
(3) Fourier-encoding in SMS MRI can also be seen as an additional 
phase-encoding in $k_z$ direction. The acquisition of many slices is therefore 
very similar to a stack-of-stars sequence in true 3D imaging for which Zhou et 
al. \cite{Zhou_Magn.Reson.Med._2017} showed that a golden angle-like rotation 
of the spoke distribution results in a higher image quality than aligned or 
linearly varied distributions.
Here the turn angle for partition $p$ is given by 
\begin{equation}
\alpha^\text{GA}_\text{trn}=\left((p-1)\cdot 
\frac{\pi}{N_\text{sp}}\cdot 
\frac{\sqrt{5}-1}{2}\right)\; \text{mod} \;\frac{\pi}{N_\text{sp}}.
\end{equation}
This scheme provides a more uniform local 3D k-space coverage as can be seen in 
Fig.\ \ref{Fig:Turn_scheme}c, but the spokes themselves are not as evenly 
distributed as in the linear-turn scheme. 
Turn-based spoke distribution schemes in 
combination with Fourier-encoded partition measurements
are known to improve image quality similar to CAIPIRINHA
in the Cartesian case \cite{Yutzy_Magn.Reson.Med._2011,Zhu__2012}.  

	\begin{figure}[htb]
		\centering
		\includegraphics[width=\textwidth]{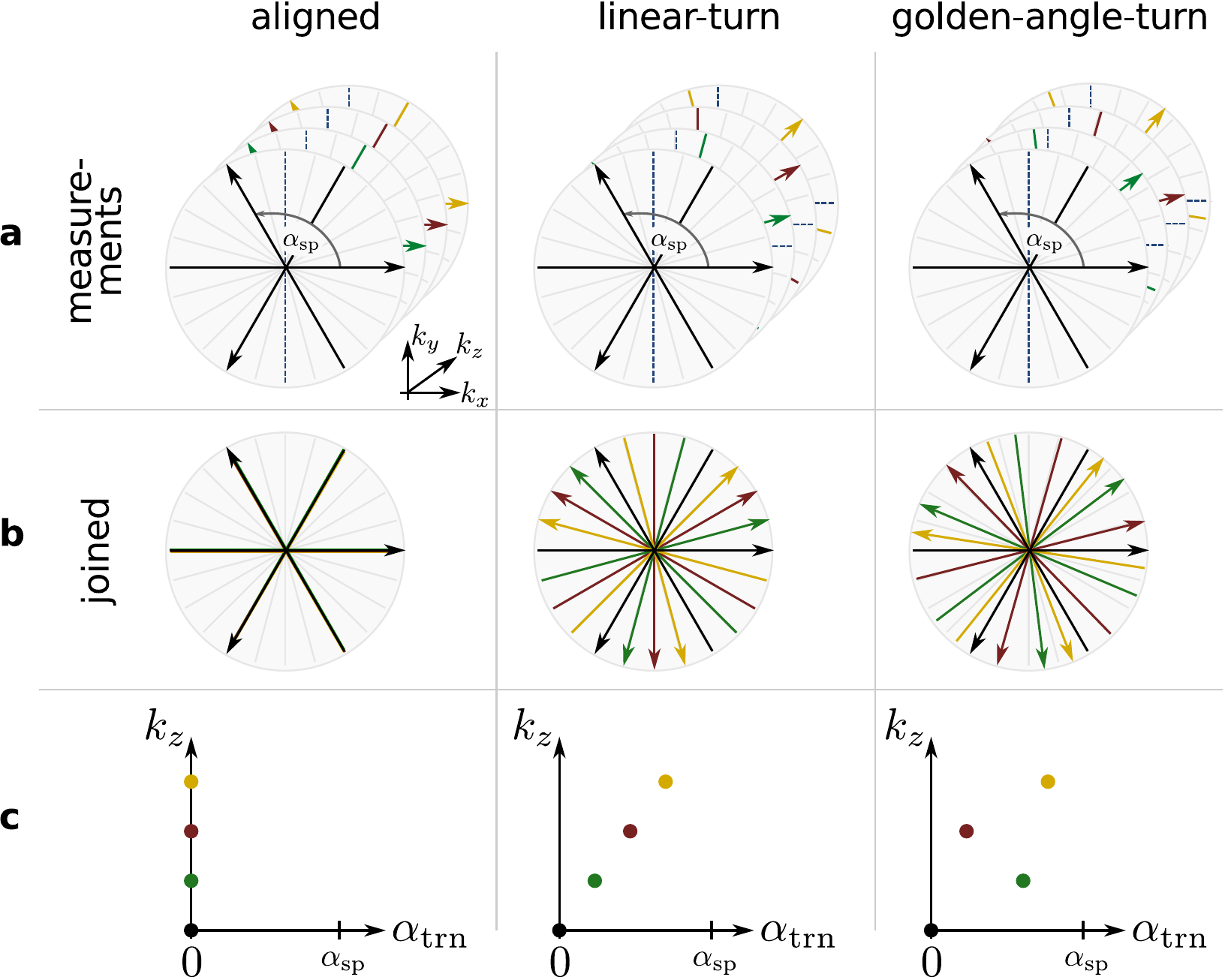}
		\caption{Schematic for three radial sampling schemes (multi-band 
			factor $M=4$ and $N_\text{sp}=3$ spokes per partition measurement). 
			Aligned: 
			Same spokes 
			acquired for each partition. Linear-turn: Linearly varied rotation 
			angle $\alpha_\text{trn}$ of the initial spoke distribution. 
			Golden-angle-turn: Rotation angle $\alpha_\text{trn}$ chosen 
			according to the 
			golden angle. 
			a) 
			Spoke distribution 
			for all 4 partitions. $\alpha_\text{sp}$ is the angle between 
			consecutive spokes. The arrow hints the readout direction. b) 
			Spokes of all partitions plotted in one 
			diagram. c) $k_z$ plotted against the rotation angle 
			$\alpha_\text{trn}$. }
		\label{Fig:Turn_scheme}
	\end{figure}

\subsection*{Post-processing}

	Although the matrix $\mathcal{W}$ promotes  
	adequate distribution of image content and coil sensitivities, the results 
	may still exhibit minor large scale 
	intensity variations compared to a conventional root-sum-of-squares (RSS) 
	reconstruction. This can be compensated for by multiplying the image 
	content with the RSS of the coil profiles:
	\begin{equation}
	m_q^\text{final} = m_q 
	\cdot \sqrt{\sum_{j=1}^N|c_q^j|^2}
	\end{equation}

\section*{Methods}

	Cartesian and radial 2D FLASH sequences with adapted RF excitation pulses for Fourier-encoded 
	SMS excitation as described in the theory section 
	were developed and utilized in this study. All experiments were
	conducted on a Magnetom Skyra 3T (Siemens Healthcare GmbH, Erlangen, 
	Germany) scanner using a 20-channel head/neck coil for phantom and human 
	brain measurements and a combined thorax and spine coil with 26 channels 
	for human heart measurements. All phantom measurements ($\text{FOV} = 
	170\times\SI{170}{\square\milli\meter}$,  
	matrix size $192 \times 192$, slice thickness $\Delta z = 
	\SI{6}{\milli\meter}$) were performed 
	on a custom-made phantom (Fig.\ \ref{Fig:Phantom}) 
	consisting of ABS bricks 
	(LEGO) being immersed in pure water.
	
	\begin{figure}[h]
		\centering
		\includegraphics[width=0.3\textwidth]{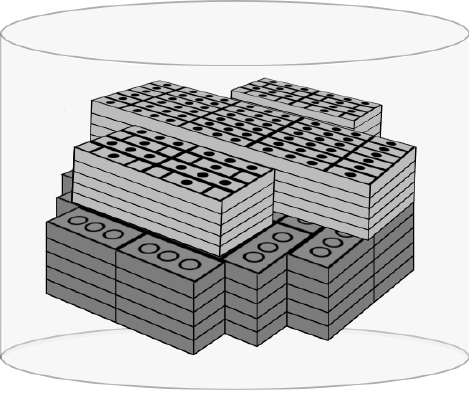}
		\caption{ Schematic of the custom-made phantom consisting of 
			LEGO bricks in pure water.}
		\label{Fig:Phantom}
	\end{figure}
	
	It is designed 
	such that the proton density of the top and bottom part of the phantom 
	differ distinctly from each other. This property can be used to demonstrate the
	capability of SMS-NLINV to disentangle simultaneously	excited slices. 
	The in-vivo brain measurements ($\text{FOV} = 
	230\times\SI{230}{\square\milli\meter}$, 
	matrix size $192 \times 192$, slice thickness $\Delta z = 
	\SI{4}{\milli\meter}$, flip angle $\theta=\SI{25}{\degree}$) as well as the 
	heart measurements ($\text{FOV} = 256\times\SI{256}{\square\milli\meter}$, 
	matrix size $160 \times 160$, slice thickness $\Delta z = 
	\SI{6}{\milli\meter}$, flip angle $\theta=\SI{8}{\degree}$) were performed 
	on volunteers with no known illnesses. 
	In all experiments, all simultaneously acquired slices are separated by 
	a fixed distance $d$.	
	NLINV as well as SMS-NLINV were implemented in the C-based 
	software package BART \cite{Uecker__2015}. The initial guess was $m_q=1$ ($ 
	q=1,\dots,M$) for the magnetizations and $c_q^j=0$ ($q=1,\dots,M,\; 
	j=1,\dots,N$) for the coil sensitivities. The parameters for the Sobolev 
	norm were set to  $a=1,\;b=220$ and $l=32$. The initial regularization 
	parameter was $\beta_0=1$ with reduction factor $h=1/2$. In the interest  
	of reproducible research, code and data to reproduce the experiments are 
	made available on 
	Github.\footnote{\url{https://github.com/mrirecon/sms-nlinv}}	

		
To confirm the basic functionality of SMS-NLINV, a Cartesian SMS 
measurement with multi-band factor $M=2$ (slice distance 
$d=\SI{60}{\milli\meter}$, TE/TR=$4.4/\allowbreak\SI{8.3}{\milli\second}$,
flip angle $\theta=\SI{15}{\degree}$) was performed on the 
brick phantom. 
A full k-space was acquired and undersampling was
performed retrospectively by multiplication with the corresponding patterns. 
The full and a retrospectively undersampled k-space (CAIPIRINHA 
pattern, $R=4$, $L_\text{ref}=12$) were reconstructed with SMS-NLINV. For 
comparison, reconstructions were also performed using the 
L2-regularized ESPIRiT 
algorithm \cite{Uecker_Magn.Reson.Med._2014}, which is based on SENSE 
\cite{Pruessmann_Magn.Reson.Med._1999} and can therefore also be applied to
SMS data \cite{Breuer_Magn.Reson.Med._2005}.
To validate the accuracy of the results, difference images between
the full and undersampled reconstructions were calculated. To assure proper 
difference images for ESPIRiT reconstructions, the complex-valued slice-images were 
multiplied with the corresponding coil sensitivities followed by a RSS 
combination. The post-processing step in SMS-NLINV already compensates for 
intensity variations, thus adequate difference images 
can be calculated using the magnitude of the resulting images. 
We performed the same
experiment using $L_\text{ref}=4$ reference lines to demonstrate the 
advantage of SMS-NLINV over ESPIRiT given a very small calibration region.

The CAIPIRINHA technique can significantly improve the image quality of SMS 
experiments \cite{Breuer_Magn.Reson.Med._2005,Zhu__2012}. We confirm
these findings for SMS-NLINV by comparing retrospectively undersampled
SMS measurements 
(TE/TR=$4.8/\SI{9.1}{\milli\second}$, flip 
angle $\theta=\SI{15}{\degree}$, $L_\text{ref}=12$, $R=4$) with CAIPIRINHA 
patterns to SMS measurements with aligned 
patterns
using the multi-band factors $M=2$ (slice distance $d=\SI{60}{\milli\meter}$) 
and $M=3$ (slice distance $d=\SI{30}{\milli\meter}$).
The absolute slice 
locations were chosen such that the outermost slices in both experiments were
located at the same positions, which allowed a comparison of the respective
slice images. A reference measurement was performed with each investigated 
slice acquired separately in a single-slice experiment and reconstructed with 
regular NLINV using equivalent reconstruction parameters. Apart from reduced 
SNR, the single-slice measurements should be identical to the acquisition with 
the aligned patterns.

The same experiment was performed using a radial trajectory to rule out errors 
with the radial SMS-FLASH sequence 
(TE/TR=$2.0/\SI{3.1}{\milli\second}$, flip angle $\theta=\SI{15}{\degree}$,
$N_\text{sp}=29$ spokes per partition) and the SMS-NLINV reconstruction for 
non-Cartesian data. 
Again, the improved k-space coverage of interleaved acquisitions, i.e. the 
use of
linear-turn- 
and golden-angle-turn-based spoke 
distributions,
should provide better results than aligned distributions or single-slice 
measurements with the same reduction factor. As a reference we performed single 
slice measurements on the same slices using $N_\text{sp}=301$ spokes to achieve 
Nyquist sampling even in the outer region of k-space. 
Finally, we present two in-vivo experiments. 
First, $M=5$ slices (slice distance 
$d=\SI{60}{\milli\meter}$, TE/TR=$4.0/\SI{9.8}{\milli\second}$, 
$N_\text{sp}=39$ spokes per 
partition) of a human brain
were acquired using the golden-angle-turn scheme. 
Reconstructions were performed
using SMS-NLINV and L2-ESPIRiT.
Calibration using ESPIRiT requires a four step procedure:
(1) the reconstruction of a fully-sampled
Cartesian calibration regions using gridding (for all partitions)
(3) disentangling of the partitions into slices using the inverse of
matrix $\Xi$,  (3) Fourier transformation back into a Cartesian 
k-space (for each slice), and (4) actual calibration from the Cartesian 
k-space data (for each slice). For this procedure to work, only the
region in k-space which fulfills the Nyquist criterion 
in all partitions can be used. The size of the
calibration region for ESPIRiT $R_\text{cal}$ is limited by 
the Nyquist criterion and was calculated to 
be $R_\text{cal} = 35\times35$. Second, $M=2$ slices (slice 
distance $d=\SI{40}{\milli\meter}$, TE/TR=$1.37/\SI{2.2}{\milli\second}$, 
$N_\text{sp}=35$ spokes per partition) of a human heart were 
acquired without ECG-triggering \cite{Uecker_NMRBiomed._2010}.
To be able to reconstruct a single frame without temporal regularization
or filtering, we combined 5 interleaves with 7 spokes per partition to 
obtain a single data set with 35 spokes per partition and linear-turn
scheme.


\section*{Results}

\subsection*{Cartesian data}

Figure \ref{Fig:SMS-NLINV_ESPIRiT}a shows ESPIRiT and SMS-NLINV 
reconstructions of a 4-fold undersampled Cartesian SMS measurement with 
multi-band factor 
$M=2$ and $L_\text{ref}=12$ reference lines. The SMS-NLINV algorithm can 
completely disentangle the superposed 
slices without significant artifacts after $it=9$ Newton steps. The 
resulting image quality is equivalent to ESPIRiT.
Figure \ref{Fig:SMS-NLINV_ESPIRiT}a also depicts difference images of 
undersampled and full reconstructions for both methods. For better visibility 
the image intensity was 
increased by a factor of 5. 
In 
all 
difference images almost no residual image content can be 
observed and mostly noise is present, which means that almost all aliasing 
artifacts
could be eliminated. The enhanced noise
in the central region is a consequence of the specific Cartesian
sampling pattern.
Figure 
\ref{Fig:SMS-NLINV_ESPIRiT}b 
shows 
the same reconstructions using a reduced calibration region. Whereas we find 
significant aliasing artifacts for ESPIRiT, SMS-NLINV still 
provides good results after $it=10$ Newton steps.

\begin{figure}[h]
		\centering
		\includegraphics[width=0.8\textwidth]{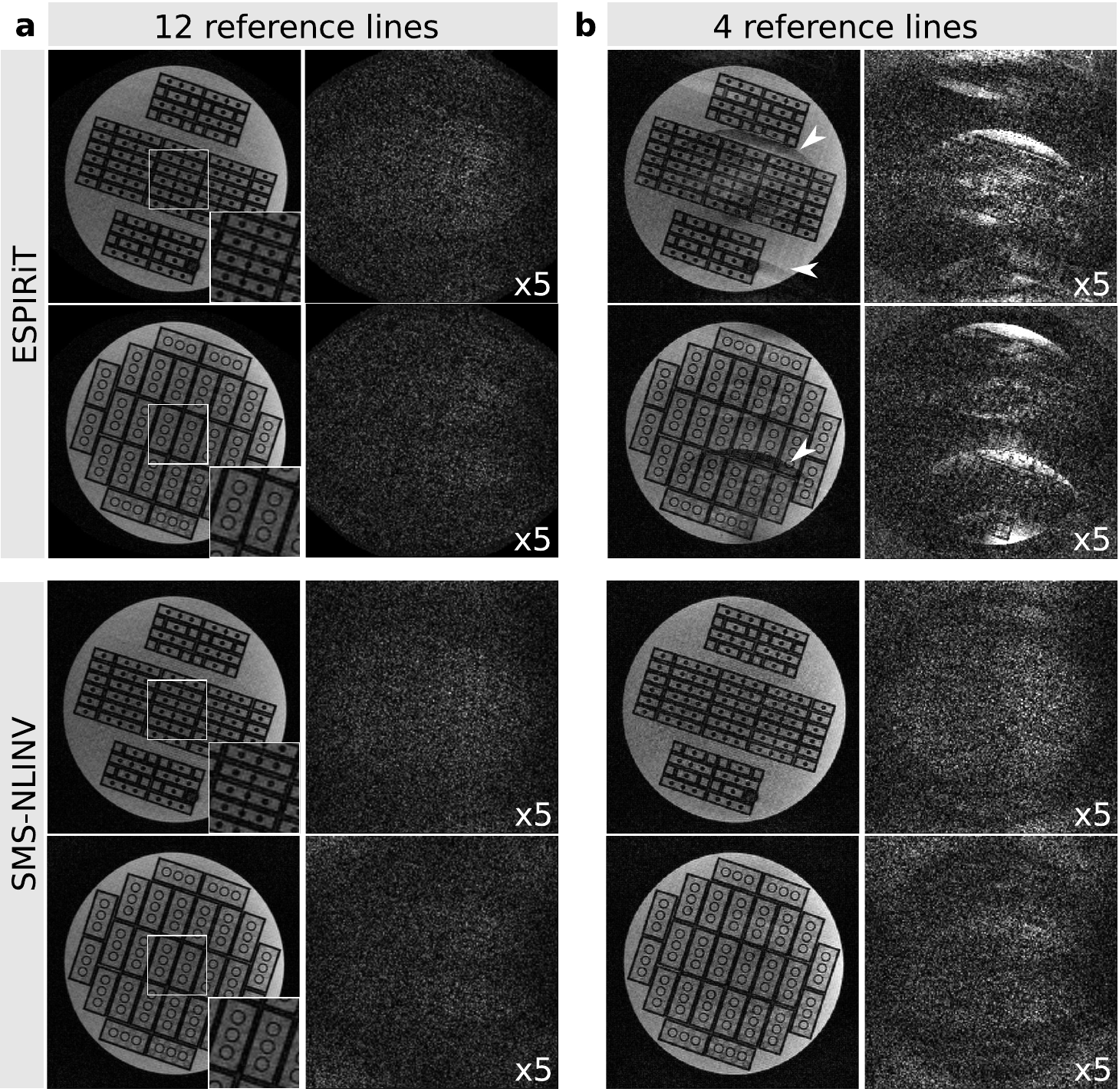}
		\caption{ a) Reconstructions of a 4-fold undersampled 
		Cartesian SMS measurement with multi-band factor $M=2$ (slice distance 
		$d=\SI{60}{\milli\meter}$, $L_\text{ref}=12$ reference lines) and 
		corresponding difference images to full 
		reconstructions: ESPIRiT and SMS-NLINV after $it=9$ 
		Newton steps.  A magnified region-of-in\textsl{}terest indicated by a 
		white rectangle is shown as inset on the 
		bottom right. For better visibility, the image intensity of the 
		difference images was increased by a factor of 5. b) Same 
		experiment as in a) using only $L_\text{ref}=4$ reference lines 
		and 
		$it=10$ Newton steps for SMS-NLINV. The arrows highlight aliasing 
		artifacts.}
		\label{Fig:SMS-NLINV_ESPIRiT}
\end{figure}

Figure \ref{Fig:Cartesian_pattern_exp} shows SMS-NLINV reconstructions of 
Cartesian SMS 
acquisitions with multi-band factors $M=2$ and $M=3$ after $it=10$ 
Newton steps using aligned and CAIPIRINHA patterns. As a 
comparison, the figure also depicts NLINV reconstructions of single-slice 
measurements for the same slices using $it=10$ Newton steps.
For both pattern types and multi-band factors the superposition can be
completely disentangled and no severe undersampling artifacts are 
present. However, the image quality resulting from the CAIPIRINHA data is 
clearly superior to aligned SMS and single-slice data. Besides an SNR benefit, 
the images of aligned SMS do not show any advantages compared to the 
single-slice images. By contrast, the CAIPIRINHA images 
resolve small details of the 
phantom bricks much better.

	\begin{figure}[h]
		\centering
		\includegraphics[width=\textwidth]{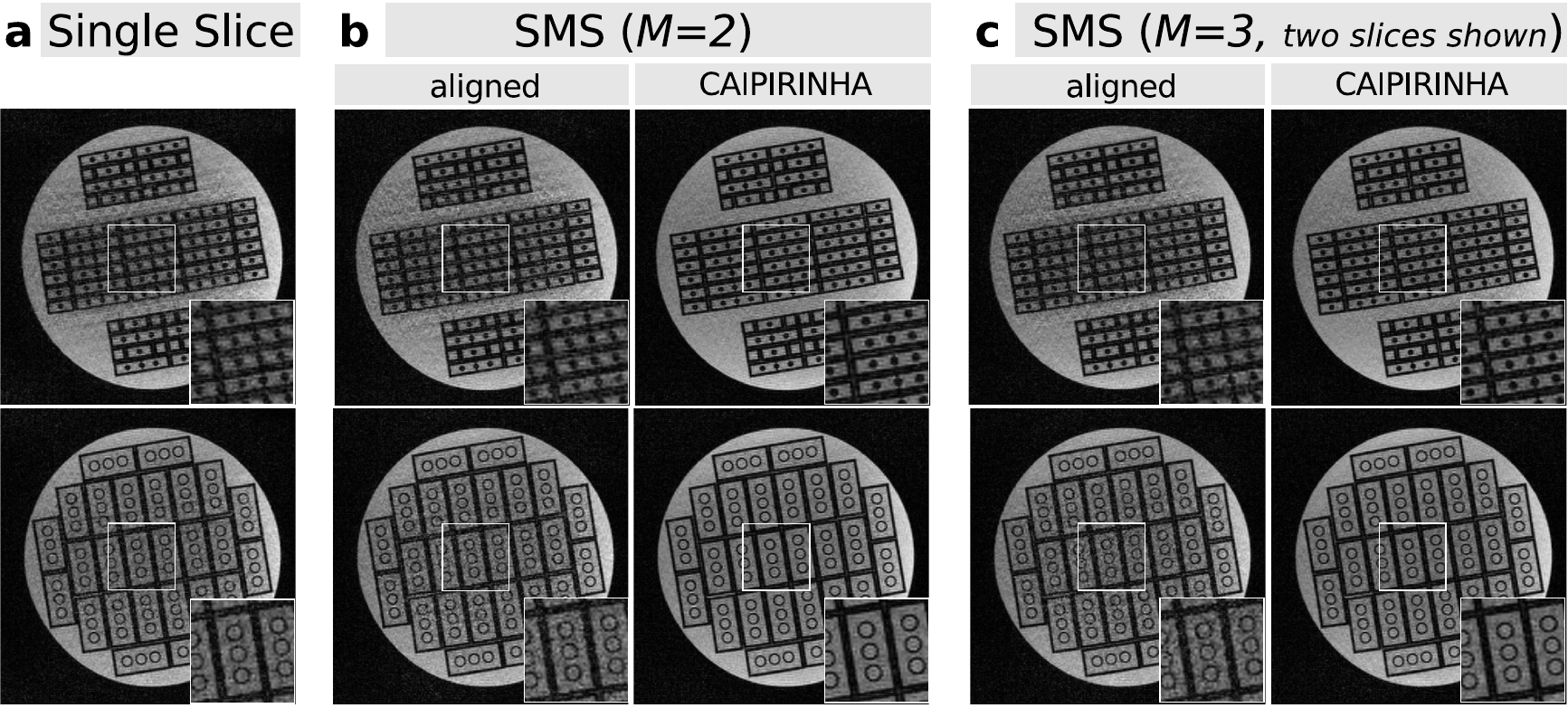}
		\caption{Comparison of different acquisition and reconstruction 
			strategies for Cartesian measurements on the brick phantom with 
			reduction factor $R=4$ and $L_\text{ref}=12$ reference lines. a) 
			Single-slice 
			acquisition and NLINV reconstruction 
			for each slice. b) SMS acquisition and SMS-NLINV reconstruction 
			for $M=2$ and aligned (left) and CAIPIRINHA pattern
			(right). Slice distance $d=\SI{60}{\milli\meter}$. c) SMS 
			acquisition and SMS-NLINV reconstruction 
			for $M=3$ and aligned (left) and CAIPIRINHA pattern
			(right). Only the outermost slices with slice distance 
			$d=\SI{60}{\milli\meter}$ are depicted. A magnified 
			region-of-interest indicated by a white rectangle is shown as inset 
			on the 
			bottom right of every image.}
		\label{Fig:Cartesian_pattern_exp}
	\end{figure}

\subsection*{Radial data}


Figure \ref{Fig:Turn_exp} depicts SMS-NLINV reconstructions of aligned, 
linear-turn- and golden-angle-turn-based radial SMS acquisitions with 
multi-band 
factor $M=3$ after 
$it=10$ 
Newton steps, as well as NLINV reconstructions of single-slice 
measurements for the same slices using $it=10$ Newton steps. 
The results 
for $M=2$ are provided as supporting Figure S1. Similar to Fig.\  
\ref{Fig:Cartesian_pattern_exp} the slice images could be reconstructed without 
significant undersampling or superposition artifacts. As in the Cartesian case, 
the turn-based SMS 
acquisitions where complementary k-space data are acquired in each partition 
yield a much better image quality than aligned SMS and 
single-slice measurements. 
The linear-turn and the golden-angle-turn scheme yield similar results. 
As supporting Figure S2 we provide difference images in image and k-space 
for $N_\text{sp}=301$ and $N_\text{sp}=29$ spokes per partition for the 
linear-turn-based $M=3$ measurement. Supporting Figure S3 shows the same 
experiments as Figure \ref{Fig:Turn_exp} and supporting Fig. S1 but with 
$N_\text{sp}=69$ spokes.

		\begin{figure}[h]
			\centering
			\includegraphics[width=\textwidth]{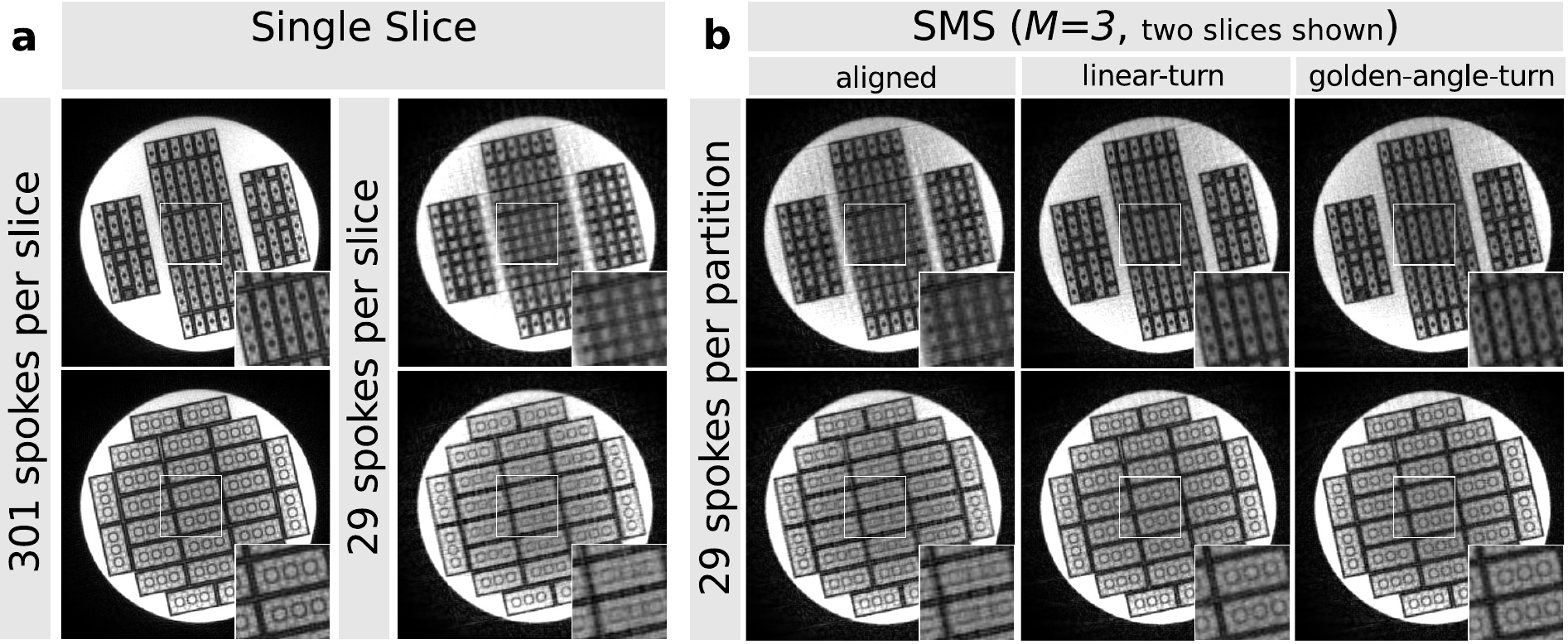}
			\caption{Comparison of different acquisition and reconstruction 
			strategies for radial measurements on the brick phantom with 
			$N_\text{sp}=29$ spokes per partition or slice and a fully sampled 
			reference scan 
			with $N_\text{sp}=301$ spokes per slice. a) 
			Single-slice 
			acquisition and NLINV reconstruction 
			for each slice. b) SMS 
			acquisition and SMS-NLINV reconstruction 
			for $M=3$ and aligned (left), linear-turn-based (center) and 
			golden-angle-turn-based sampling (right). Only the outermost slices 
			with slice distance 
			$d=\SI{60}{\milli\meter}$ are depicted. A magnified 
			region-of-interest indicated by a white rectangle is shown as inset 
			on the 
			bottom right of every image. The same experiment for $M=2$ is 
			provided as supporting Figure S1.}
			\label{Fig:Turn_exp}
	\end{figure}


Figure \ref{Fig:in vivo_brain} and \ref{Fig:in vivo_heart} show the results of 
the in-vivo scans where we 
have 
chosen the number of Newton steps to obtain the best results using visual 
observation. Figure \ref{Fig:in 
vivo_brain} shows $M=5$ slices of a 7-fold undersampled 
acquisition 
of a human brain reconstructed with SMS-NLINV using $it=11$ Newton steps and 
ESPIRiT. Both methods show similar results as all 
slices are completely disentangled and all streaking artifacts could be 
eliminated.
As supporting Figure S4 shows, the residual for the SMS-NLINV reconstruction
approaches a 
constant value when plotted against the number of Newton steps.

		\begin{figure}[h]
			\centering
			\includegraphics[width=\textwidth]{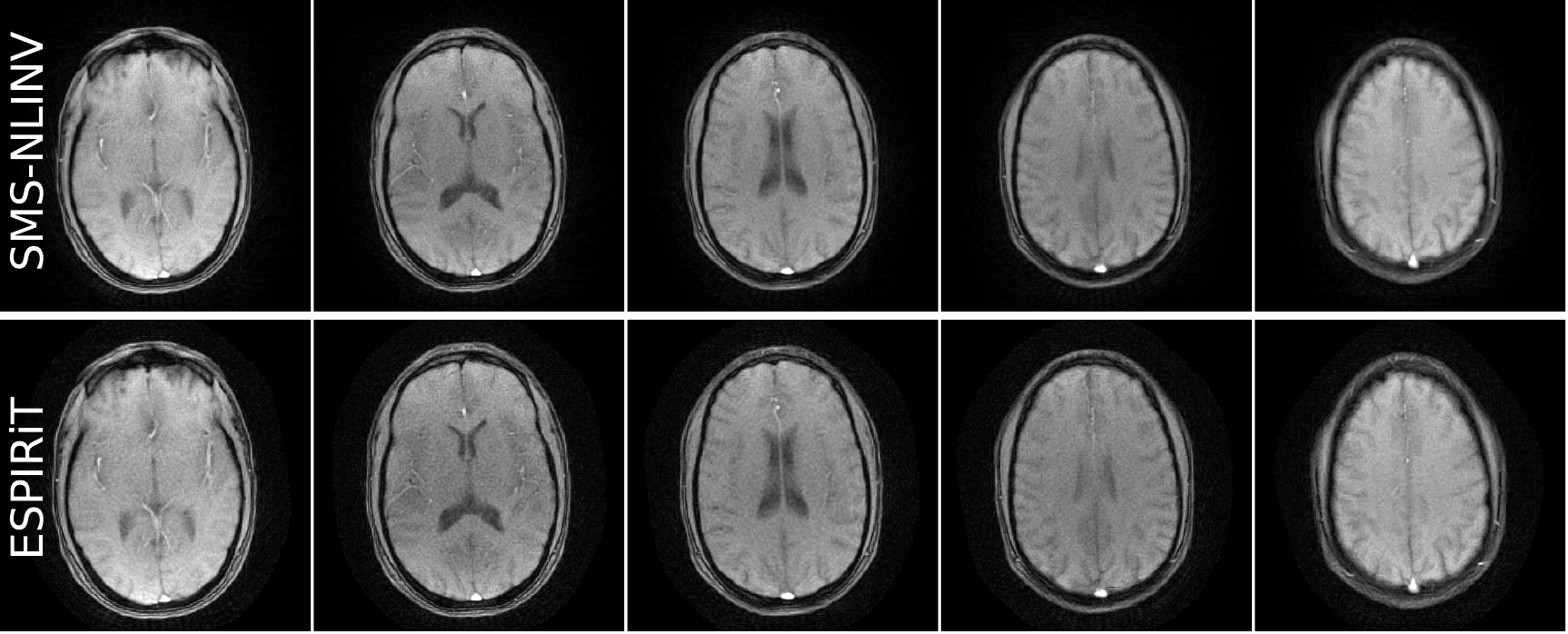}
			\caption{SMS-NLINV and ESPIRiT 
			reconstructions 
			of a 
			human brain using
			 a radial SMS-FLASH acquisition with multi-band factor $M=5$
			(slice distance 
			$d=\SI{10}{\milli\meter}$, slice thickness $\Delta z = 
			\SI{4}{\milli\meter}$, golden-angle-turn scheme, 
			$N_\text{sp}=39$ spokes per partition,
			$it=11$ Newton steps).}
			\label{Fig:in vivo_brain}
		\end{figure}

Figure \ref{Fig:in vivo_heart} depicts $M=2$ slices of a human 
heart in end-diastole simultaneously acquired in  $\SI{154}{\milli\second}$ 
using 
$N_\text{sp}=35$ spokes per partition and  
reconstructed with SMS-NLINV using $it=13$ Newton steps.
Again, the two slices 
are completely disentangled and only slight blurring as well as 
minor streaking artifacts are present.  

	\begin{figure}[h]
		\centering
		\includegraphics[width=\textwidth]{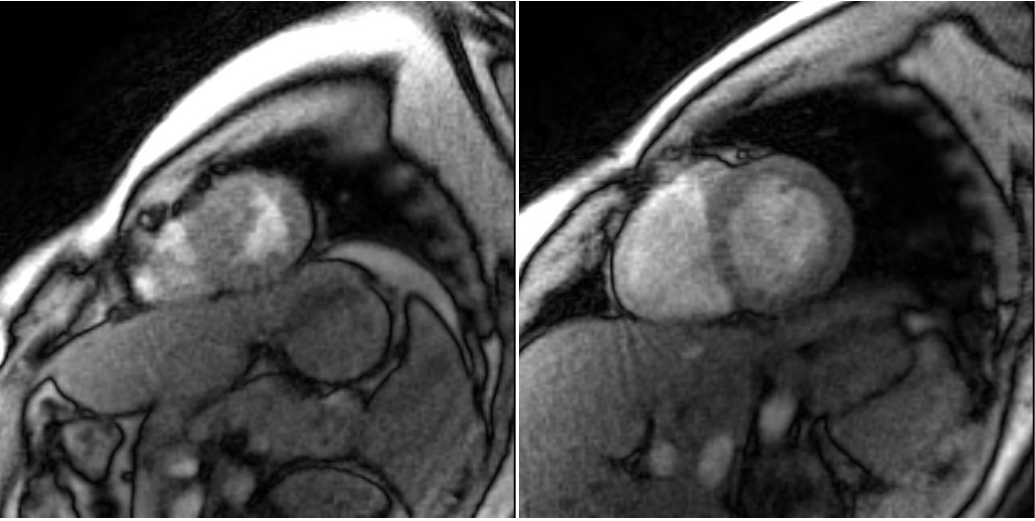}
		\caption{SMS-NLINV reconstruction of a 
			human 
			heart 
			in end-diastole 
			using a real-time SMS-FLASH acquisition with 
			multi-band factor $M=2$ (slice distance 
			$d=\SI{40}{\milli\meter}$, slice thickness $\Delta z = 
			\SI{6}{\milli\meter}$, $N_\text{sp}=35$ spokes per partition,
			$it=13$ Newton steps).}
		\label{Fig:in vivo_heart}
	\end{figure}


\section*{Discussion}

In this work, SMS-NLINV has been combined with
a Cartesian and a radial SMS-FLASH sequence.


With sufficient reference lines, SMS-NLINV and
ESPIRiT reconstruct undersampled SMS data with similar image quality as
shown in this work for radial and Cartesian data. This
finding is in agreement with previous results comparing ESPIRiT and
regular NLINV \cite{Uecker_Magn.Reson.Med._2014}.
For very small calibration regions 
SMS-NLINV still provides good results where ESPIRiT reconstructions already 
show severe artifacts. The reason for this is that SMS-NLINV does not rely on
a fully sampled calibration region but jointly estimates the image content and
the coil sensitivities. In contrast, direct calibration using ESPIRiT or other
calibration methods requires a complicated four step procedure
and only the region in k-space which fulfills the 
Nyquist criterion simultaneously in all partitions can be utilized for
calibration. SMS-NLINV not only makes all these additional
separate processing steps unnecessary, it also works even for very 
small calibration regions by exploiting all available samples. 
This latter property makes SMS-NLINV ideally suited for non-Cartesian
sampling, especially for accelerated dynamic imaging with changing coil 
sensitivities where only 5 to 7 spokes per partition are acquired and the
Nyquist-sampled region can become very small \cite{Rosenzweig__2017}.
We compared Cartesian SMS acquisitions with CAIPIRINHA and 
aligned patterns as well as radial SMS acquisitions with aligned spoke
scheme and linear-turn spoke schemes for different multi-band factors $M$
with single-slice ($M=1$) measurements as control.
The aligned schemes acquire the same k-space samples for each of the $M$ 
partitions which resembles an averaging process and thus yields an SNR 
benefit relative to single-slice measurements. 
However, the partitions do not contain complementary 
k-space information and therefore no significant advantages in terms of better 
resolved details can be achieved. In this case a joint reconstruction
does not possess any benefits compared to an 
inverse discrete Fourier transform (Eq.\ (\ref{Eq:k encoded inverse})) on the 
Fourier-encoded k-spaces followed by single-slice reconstructions.
The actual advantage of SMS-NLINV becomes apparent with the use of schemes 
where each partition contributes complementary k-space 
data which is equivalent to supplementary object information. 
Consequently, in addition to the SNR benefit, details are 
better resolved. Whereas single-slice NLINV and aligned SMS-NLINV 
recover missing k-space samples using 2D sensitivity information, 
SMS-NLINV using complementary data also exploits sensitivity variations in the 
third 
dimension which allows for higher acceleration factors 
\cite{Weiger_Magn.Reson.Mater.Phys.Biol.Med._2002}. 

In principle, the more slices $M$ we simultaneously 
acquire using an interleaved scheme, the better will be the resulting image 
quality of each slice due to the $\sqrt{M}$-like SNR benefit and the 
acquisition of additional complementary samples. 
	However, for the acquisition of $M$ slices we have to perform $M$ partition 
	measurements and therefore the measurement time increases with increasing 
	$M$ until it approaches the time of a 3D measurement. The 
	optimal 
	choice for $M$ depends on various experimental considerations such as 
	overall motion robustness,
	magnetization 
	preparation scheme, etc.

A future subject of study will be the use of SMS-NLINV for 
dynamic imaging 
at high temporal 
resolution, 
which was already successfully demonstrated for single-slice MRI using NLINV 
\cite{Uecker_Magn.Reson.Med._2010, Uecker_NMRBiomed._2010, 
Voit_J.Cardiov.Magn.Reson._2013, Zhang_J.Cardiov.Magn.Reson._2010}. 
The 
incorporation of additional minimization penalties such as temporal 
regularization and median filtering \cite{Uecker_Magn.Reson.Med._2010} known 
from NLINV and adapted 
spoke distribution schemes for dynamic SMS imaging 
\cite{Wu_J.Magn.Reson.Imaging_2016} can directly be applied to SMS-NLINV and 
will further reduce 
streaking artifacts as well as blurring and improve the overall 
image quality. Preliminary 
results have been presented by Rosenzweig et al.\ \cite{Rosenzweig__2017}.


In this work, we used a basic SMS-FLASH sequence.
However, SMS-NLINV is a very general reconstruction approach and
should be applicable to all sequences that can make use of
an SMS acquisition, such as diffusion tensor imaging
(DTI), functional MRI (fMRI) or $T_1$/$T_2$ quantification.
In the future, we also plan to combine SMS-NLINV with a bSSFP 
sequence \cite{Staeb_Magn.Reson.Med._2011} and 
more advanced regularization techniques which will improve 
image quality at high acceleration
\cite{Uecker__2008,Knoll_Magn.Reson.Med._2011}.

\section*{Conclusion}

The present work extends the NLINV algorithm to simultaneous multi-slice MRI.
As NLINV does not rely on the presence of Cartesian calibration data, it is
an ideal choice for parallel imaging with non-Cartesian acquisitions.
The combination with simultaneous multi-slice (SMS) offers the advantages of 
increased SNR
and higher acceleration by exploiting three-dimensional sensitivity encoding.

\appendix
\section*{Appendix}
\subsection*{Glossary}
\begin{table}[h]
	\caption{Glossary of notations}
	\centering
	\begin{tabular}{cl}
		\hline
		$M$ & Multi-band factor \\
		$N$ & Number of receive channels \\
		$d$ & Slice distance  \\
		${\bm{x}_q}$ & Magnetization and coil sensitivities for slice $q$\\
		${\bm{y}}$ & k-spaces of all coils\\ 
		${y}^j$ & k-space of coil $j$\\
		${y}_q$ & k-space of slice $q$  \\
		$\bm{m}$ & Magnetizations seen by all coils\\  
		$m^j$ & Magnetization seen by coil $j$\\
		$m_q$ & Magnetization of slice $q$\\
		$\bm{c}$ & Coil sensitivities of all coils\\   
		$c_q$ & Coil sensitivity of slice $q$\\
		$\hat{c}$ 	& Normalized coil sensitivity\\
		$z_q$ & Center coordinate of slice $q$ \\
		$\Delta z$ & Slice thickness\\
		$\tilde{}$ & Encoded quantity \\
		$^H$ & Adjoint \\
		$^T$ & Transpose\\
		$^*$& Complex conjugate \\
		$it$ & Number of Newton steps.\\
		\hline
	\end{tabular}
	\label{Tab:Notation}
\end{table}

\subsection*{Derivative and adjoint of the Forward operator}
\label{App:Derivatives}
Given the forward operator $F(x)$ from Eq.\ (\ref{Eq:Forward op}) the 
corresponding 
derivative reads
\begin{equation}
\label{eq:Derivative}
DF\big|_X\left(\begin{array}{c}
d\bm{x}_1\\
\vdots \\
d\bm{x}_{M}
\end{array}\right)= 		
\bm{P}
\Xi
\left(\begin{array}{c}
\mathcal{F}(dm_1 \bm{c}_1 + m_1 d\bm{c}_1) \\
\vdots \\
\mathcal{F}(dm_{M} \bm{c}_{M} + m_{M} d\bm{c}_{M})
\end{array}\right).
\end{equation}
The adjoint of the derivative is given by
\begin{gather}
DF^H\big|_X\left(\begin{array}{c}
d\tilde{\bm{y}}_1\\
\vdots \\
d\tilde{\bm{y}}_{M}
\end{array}\right) = 
\left(\begin{array}{ccc}
\left(\begin{array}{c}
\bm{c}^H_1\\
m_1^{H}		
\end{array}\right) & & 0\\
& \ddots &\\
0& &\left(\begin{array}{c}
\bm{c}_{M}^{H}\\
m_{M}^{H}		
\end{array}\right) 
\end{array}\right)
\mathcal{F}^{H}
\Xi^H
\bm{P}^H
\left(\begin{array}{c}
d\tilde{\bm{y}}_1\\
\vdots\\
d\tilde{\bm{y}}_{M}	
\end{array}\right),
\label{eq:Adjoint}
\end{gather}
with 
\begin{equation}
\nonumber 
\left(\begin{array}{c}
\bm{c}_{q}^{H}\\
m_{q}^{H}		
\end{array}\right) := 
\left(\begin{array}{c}
{c_q^{1}}^*,\dots,{c_{q}^{N}}^*\\
m_q^{*}		
\end{array}\right).
\end{equation}
The asterisk $*$ denotes pointwise complex conjugation.

\subsection*{Weighting Matrix}

The weighting matrix used in SMS-NLINV to implement the smoothness penalty
for the coil sensitivities is given by:
\label{App:Weights}
	\begin{equation}
	\mathcal{W}^{-1} := \left(\begin{array}{ccc}
	W^{-1} & & 0 \\
	&\ddots&\\
	0 &  & W^{-1}
	\end{array}\right),
	\end{equation}
Here, $W^{-1}$ is the same weighting matrix as used in conventional NLINV:
	\begin{equation}
	W^{-1}:=\left(\begin{array}{cccc}
	I & & & 0\\
	& a(1+b||\vec{k}||^2)^{l/2}\mathcal{F} & & \\
	&  & \ddots & \\
	0& & & a(1+b||\vec{k}||^2)^{l/2}\mathcal{F}
	\end{array}\right).
	\end{equation}

\newpage
\bibliographystyle{mrm}
\bibliography{radiology}

\newpage

\begin{figure}[h]
	\centering
	\includegraphics[width=\textwidth]{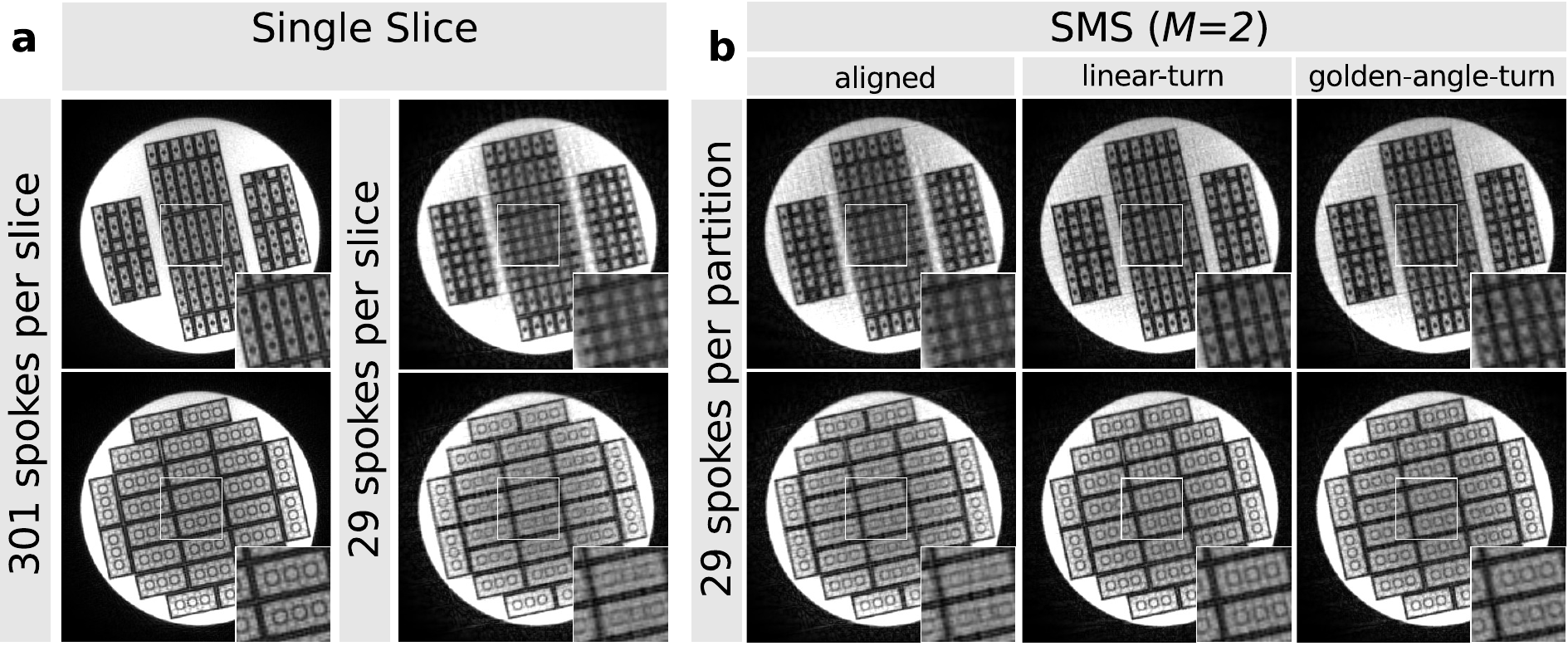}
	\caption*{Supporting Figure S1: Comparison of different acquisition and 
reconstruction 
strategies for radial measurements on the brick phantom with 
$N_\text{sp}=29$ spokes per partition or slice and a fully sampled 
reference scan 
with $N_\text{sp}=301$ spokes per slice. a) 
Single-slice 
acquisition and NLINV reconstruction 
for each slice. b) SMS acquisition and SMS-NLINV reconstruction 
for $M=2$ and aligned (left), linear-turn-based
(center) and golden-angle-turn-based sampling (right). Slice 
distance $d=\SI{60}{\milli\meter}$. A magnified 
region-of-interest indicated by a white rectangle is shown as inset 
on the 
bottom right of every image.}
\end{figure}

\begin{figure}[h]
	\centering
	\includegraphics[width=0.6\textwidth]{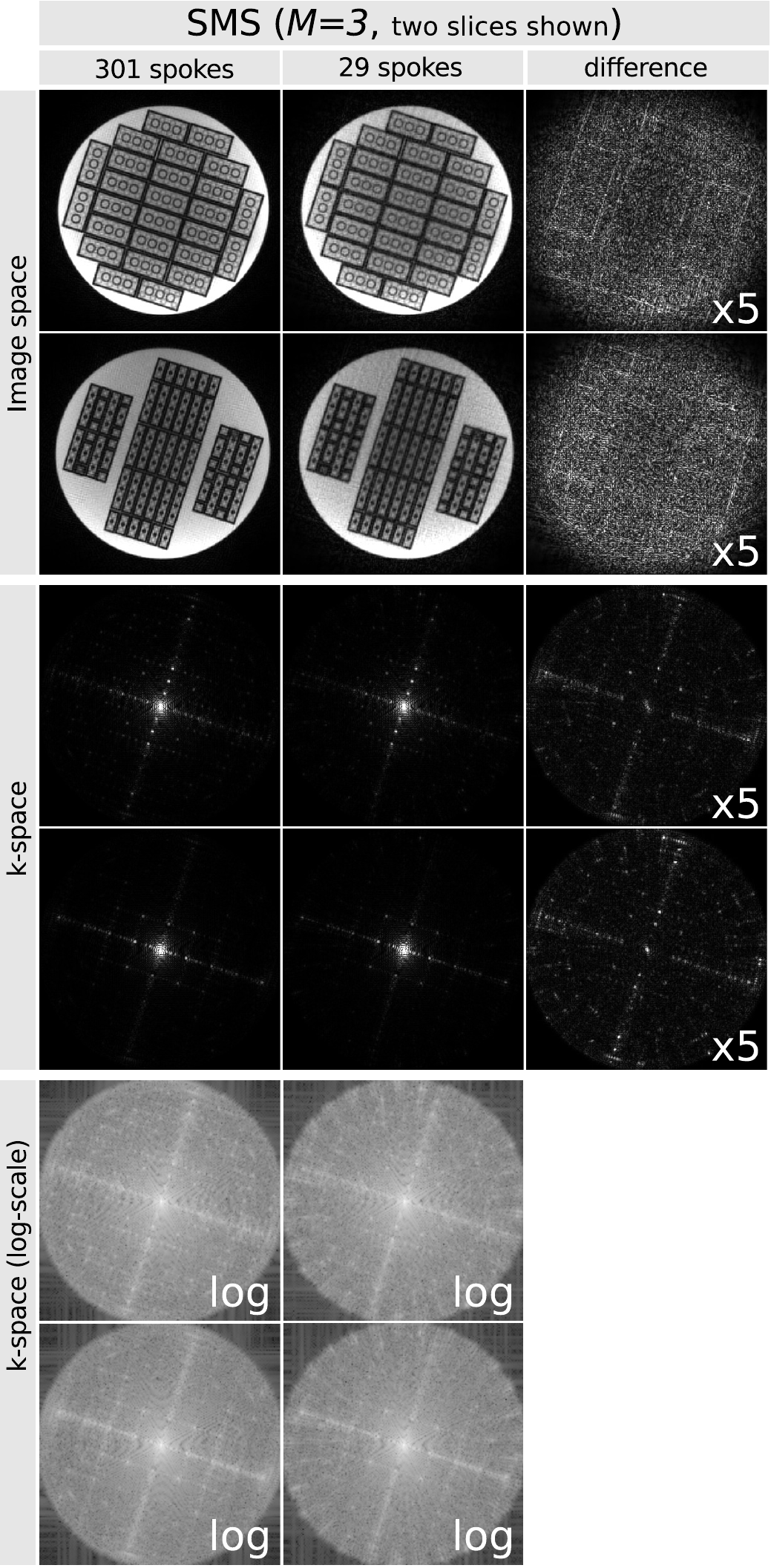}
	\caption*{Supporting Figure S2: Difference images in 
		image and k-space for 
		SMS ($M=3$, slice distance $d=\SI{30}{\milli\meter}$, 
		linear-turn-based spoke 
		distribution) acquisitions with 
		$N_\text{sp}=301$ (fully sampled reference) and $N_\text{sp}=29$ 
		spokes per 
		partition. For better visibility, the intensity of the difference 
		images was 
		increased by a factor of 5 and the k-spaces were additionally 
		depicted using the log-scale.}
\end{figure}

\begin{figure}[h]
	\centering
	\includegraphics[width=0.5\textwidth]{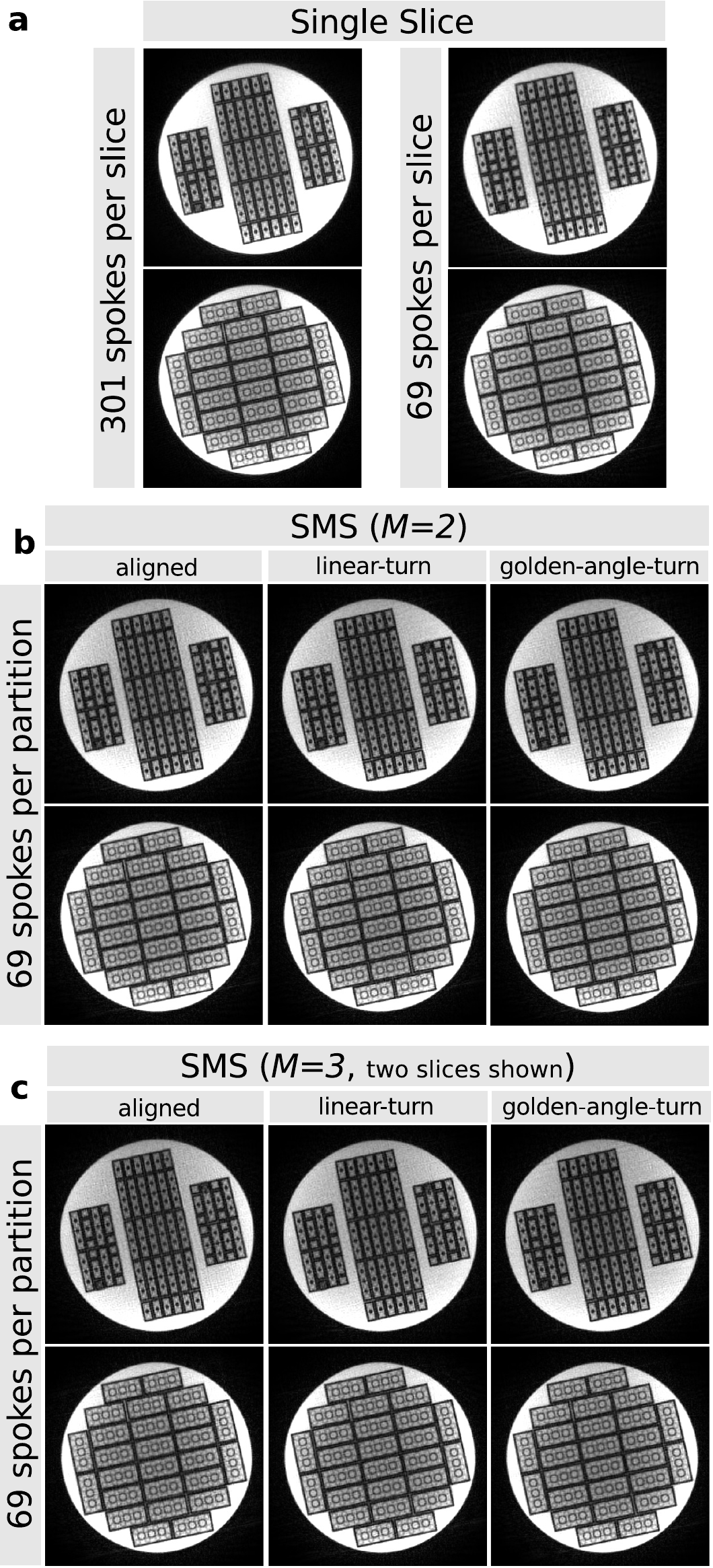}
	\caption*{Supporting Figure S3: Comparison of 
		different 
		acquisition and reconstruction 
		strategies for radial measurements on the brick phantom with 
		$N_\text{sp}=69$ spokes per partition or slice and a fully 
		sampled 
		reference scan 
		with $N_\text{sp}=301$ spokes per slice. a) 
		Single-slice 
		acquisition and NLINV reconstruction 
		for each slice. b) SMS acquisition and SMS-NLINV reconstruction 
		for $M=2$ and aligned (left), linear-turn-based
		(center) and golden-angle-turn-based sampling (right). Slice 
		distance $d=\SI{60}{\milli\meter}$. c) SMS 
		acquisition and SMS-NLINV reconstruction 
		for $M=3$ and aligned (left), linear-turn-based (center) and 
		golden-angle-turn-based sampling (right). Only the outermost 
		slices 
		with slice distance 
		$d=\SI{60}{\milli\meter}$ are depicted.}
\end{figure}

\begin{figure}[h]
	\centering
	\includegraphics[width=0.8\textwidth]{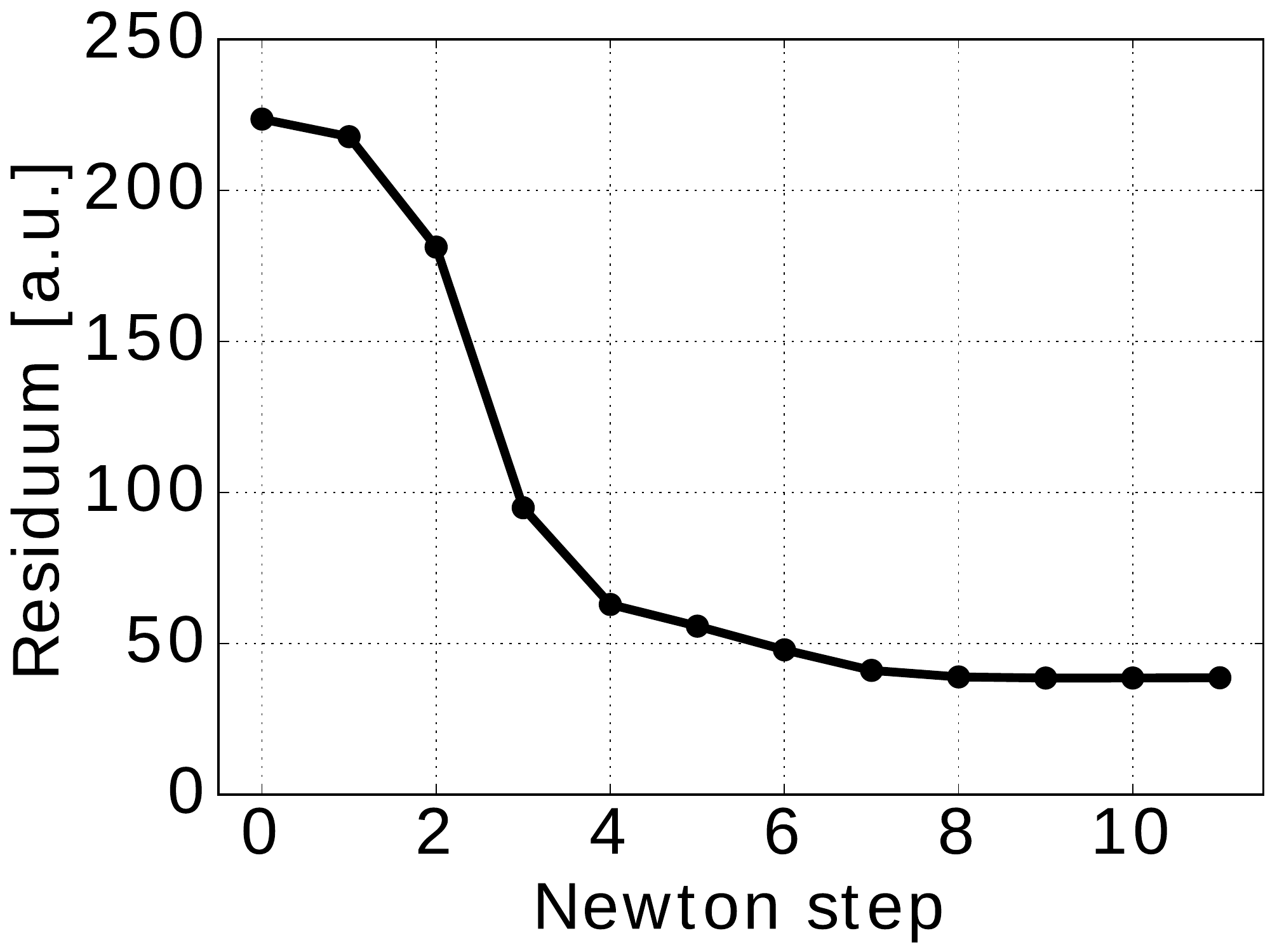}
	\caption*{Supporting Figure S4: Supporting Figure S4: Residuum of the 
		SMS-NLINV reconstruction in 
		Figure 7 
		against the number of Newton steps.}
\end{figure}

\end{document}